# Environmental sound analysis with mixup based multitask learning and cross-task fusion

Weiping Zheng, Dacan Jiang, Gansen Zhao

*Abstract*—**Environmental sound analysis is currently getting more and more attentions. In the domain, acoustic scene classification and acoustic event classification are two closely related tasks. In this letter, a two-stage method is proposed for the above tasks. In the first stage, a mixup based MTL solution is proposed to classify both tasks in one single convolutional neural network. Artificial multi-label samples are used in the training of the MTL model, which are mixed up using existing single-task datasets. The multi-task model obtained can effectively recognize both the acoustic scenes and events. Compared with other methods such as re-annotation or synthesis, the mixup based MTL is low-cost, flexible and effective. In the second stage, the MTL model is modified into a single-task model which is fine-tuned using the original dataset corresponding to the specific task. By controlling the frozen layers carefully, the task-specific high level features are fused and the performance of the single classification task is further improved. The proposed method has confirmed the complementary characteristics of acoustic scene and acoustic event classifications. Finally, enhanced by ensemble learning, a satisfactory accuracy of 84.5 percent on TUT acoustic scene 2017 dataset and an accuracy of 77.5 percent on ESC-50 dataset are achieved respectively.**

## I. INTRODUCTION

IN recent years growing interests have been shown towards automatic analysis of environmental sounds [1]. The ability of understanding surrounding environment will greatly enhance the intelligence of practical applications, such as campus security and protection, robot navigation, etc. In the domain, there exist two main tasks: acoustic scene classification (ASC) and acoustic event detection (AED). Acoustic scene refers to the entirety of sound that is formed by various sound objects

This work was partially supported by the National Key Research and Development Program under Grant 2019YFB1804003 and Characteristic Innovation Projects of the Educational Commission of Guangdong Province, China under Grant 2016KTSCX025 and Guangzhou Science and Technology Fund under Grant 201804010314. (*Corresponding author: Weiping Zheng)

Weiping Zheng, Dacan Jiang, Gansen Zhao are with the School of Computer, South China Normal University, Guangzhou 510631, China (e-mail: zhengweiping@m.scnu.edu.cn; dcjiang@m.scnu.edu.cn; gzhao@m.scnu.edu.cn).

or events [1]. It is a natural idea to utilize characteristics of acoustic events taking place in the scene when performing ASC. For example, Cai et al. [2] proposed a Bayesian network-based approach to classify auditory context by taking information of key audio effects into account. Heittola et al. [3] developed a histogram-of-events approach for audio scene classification. Imoto et al. [4][5][6] had introduced a theory of acoustic topic for ASC. In the works, they represented an acoustic signal as a multinomial distribution over acoustic topics and an acoustic topic as a distribution over acoustic events [6]. Similarly, acoustic scene information would help in the AED task. Mesaros et al. [7] applied probabilistic latent semantic analysis to model co-occurrence of events from acoustic scenes and integrated event priors into the acoustic event detection system. Heittola et al. [8] proposed a two-stage approach for AED. According to the acoustic context recognized, specific set of acoustic event classes was selected and context-dependent acoustic model was utilized in the sound event detection stage.

More recently, multitask learning (MTL) based approaches have been proposed to perform joint analysis of acoustic scene and event [9][10][11]. It was Bear et al. [9] that first proposed joint analysis of ASC and AED. To implement MTL, they created a new dataset with both sound scene and sound event labels by synthesizing foreground events with background scenes. In the network of [9], all layers were shared by both tasks. Similarly, Tonami et al. [10] proposed an MTL based solution for joint analysis of acoustic events and scenes. In their network, except for the shared layers, scene layers and event layers were separated into two branches. Imoto et al. [11] proposed an AED method by MTL of sound events and scenes with soft labels obtained by the teacher-student learning framework [12]. The dataset used in both [10] and [11] was composed of parts of the TUT sound events 2016, 2017 and TUT acoustic scenes 2017 datasets. Manual annotation was further performed to ensure both event and scene labels available for the samples in the dataset.

Using MTL paradigm, joint environmental sound analysis is more efficient as multiple tasks can be predicted in one time by one single model. Furthermore, joint analysis will result in more promising performances for the closely related sound analysis tasks, as reported in [10], because joint learning will reduce the chances of over-fitting. However, joint learning methods suffer from the preparation of multiple labels for the samples. It is time-consuming and cost-intensive regardless of re-annotation upon existing real-world dataset [10][11] or synthesis of artificial sound clips [9].

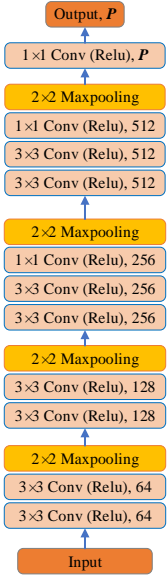

Fig.1. Baseline model. *P* equals to 15 for ASC and 50 for AEC.

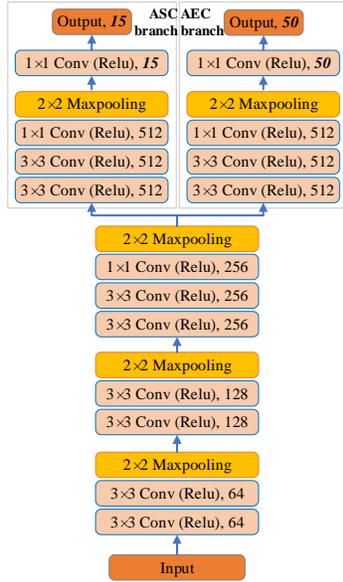

Fig. 2. MTL model for ASC and AEC tasks

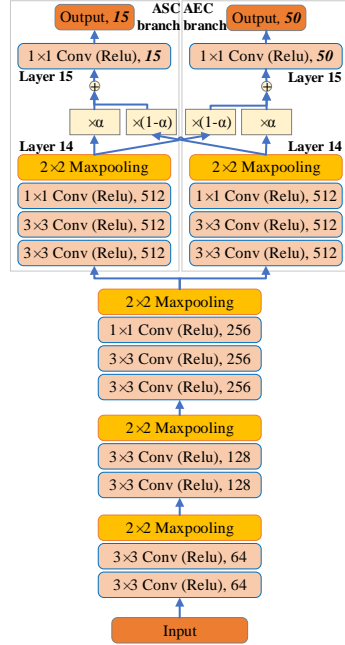

Fig. 3. MTL model with inter-connections for ASC and AEC tasks

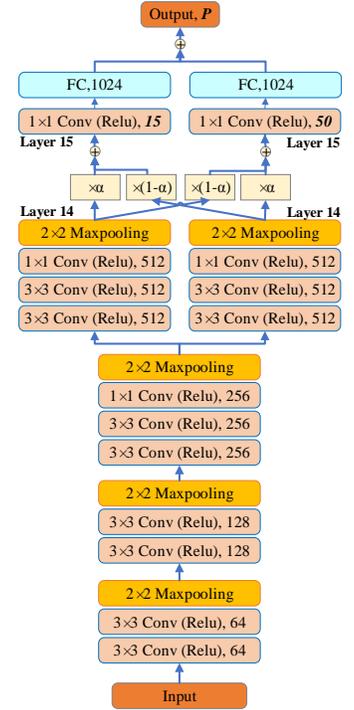

Fig. 4. Fine-tuned model. *P* equals to 15 for ASC and 50 for AEC.

In this letter, a mixup based multitask learning approach is proposed for environmental sound analysis. In the approach, only single-task datasets are needed and neither re-annotation nor synthesis is necessary. Mixup [13] is an interesting data augmentation method proposed in 2017. It constructs new training example by the linear interpolation of two random examples from the training set and their labels. In our approach, the audio signal is transformed into spectrograms. Spectrograms from different single-task datasets are mixed up into one spectrogram which is resultantly related to multiple labels. With the help of these new spectrograms, the MTL based CNN model is implemented for environmental sound analysis. To the best of our knowledge, this is the first attempt that mixup is applied to solve the multi-label construction problem in multitask learning. The main contributions of this letter are threefold:

1) We propose a mixup based MTL method for environmental sound analysis, which supplies an efficient way to construct multi-label samples by existing datasets.

2) We propose a two-stage method for environmental sound analysis. In the first stage, a multiple-branches CNN network is trained using the mixup samples. In the second stage, high-level features on each branch are fused by fine-tuning to improve the performance further.

3) Our method has achieved a satisfactory accuracy on the TUT acoustic scene 2017 dataset.

## II. PROPOSED METHOD

In this letter, we perform acoustic scene classification and acoustic event classification (AEC) on TUT acoustic scenes 2017 dataset [14] and ESC-50 dataset [15] respectively. Acoustic event classification here only predicts which event occurs in the audio segment. In this section, we first introduce a baseline model and then evolve it step by step into the final model.

### A. Baseline Model

CNN architecture has been widely adopted in ASC [16][17] and AED [18][19]. We also employ a VGGish network as the baseline model. The network structure is shown in Fig. 1. Node number and convolution kernel number in the last two layers are equal to P, which is 15 for ASC and 50 for AEC. Log-mel-spectrograms are used as input of the model. The input dimension is 128 bands ×128 frames ×1 channel. Specifically, log mel-scaled filter-bank energies are calculated using 128 mel bands, a Hann window of 2048 points and a hop length of 1024 points. Spectrogram image is generated by concatenating 128 frames and a hop step of 32 frames. To utilize the binaural information in TUT acoustic scene 2017 dataset, spectrogram images are generated from both left and right audio channels. Besides the left-channel and right-channel images, the sum of the two images is generated as another sample. These samples are used separately in the training and testing.

### B. Mixup based Multitask Learning

Multitask learning is a promising learning paradigm that is able to prevent over-fitting and improve generalization of model with the help of other related tasks [20]. However, the application of MTL is restricted owing to lack of multi-label instances. We have proposed a mixup based method to solve this problem.

Suppose $x_i^S$ is an i-th spectrogram image selected from the TUT acoustic scene 2017 training (validation) set; $y_i^S$ is the one-hot label corresponding to $x_i^S$; Accordingly, a spectrogram image $x_j^E$ is randomly selected from the ESC-50 training (validation) set; $y_j^E$ is the one-hot label corresponding to $x_j^E$; Then an i-th new sample $(\tilde{x}_i, < \tilde{y}_i^S, \tilde{y}_i^E >)$ could be constructed as below:

$$\tilde{x}_i = 0.5x_i^S + 0.5x_j^E \qquad (1)$$
$$\tilde{y}_i^S = y_i^S \qquad (2)$$
$$\tilde{y}_i^E = y_j^E \qquad (3)$$

To implement joint learning of ASC and AEC, we have constructed an MTL model based on the baseline model. The previous ten layers of the baseline model, they are seven convolution layers and three pooling layers, are extracted as the shared layers of the MTL model. Then the rest layers are duplicated. They are both connected to the shared layers. As a result, the new model has two branches, which are responsible for ASC and AEC respectively. The network structure is shown in Fig. 2, which is similar to the one in [10].

### C. Adding Inter-connections between Branches

The activation maps on each branch are considered as task-specific representations. It is expected that the representation for one task will help the classification of the other task. To this end, we simply set up cross connections between the two branches, see Fig. 3. In this way, the task of ASC will take advantage of the AEC specific representations, and vice versa. Specifically, an inter-connection unit is inserted between the $14^{th}$ layer and the $15^{th}$ layer of both branches which is a little similar with the cross-stitch unit [21].

Let $H_i^S$ and $H_i^E$ denote the output of the i-th layers on the ASC and AEC branches respectively. The inter-connection of hidden layers can be expressed as below:

$$H_{15}^S = f(W_{15}^S(\alpha \cdot H_{14}^S + (1 - \alpha) \cdot H_{14}^E) + b_{15}^S) \qquad (4)$$
$$H_{15}^E = f(W_{15}^E(\alpha \cdot H_{14}^E + (1 - \alpha) \cdot H_{14}^S) + b_{15}^E) \qquad (5)$$

where $W_{15}^S$, $W_{15}^E$ and $b_{15}^S$, $b_{15}^E$ are trainable weights and biases for the $15^{th}$ layers of each branch respectively; $f(\cdot)$ is the activation function Relu; $\alpha$ is a hyper-parameter, $0 < \alpha < 1$.

### D. Improving Single-task Performance through Cross-task Fusion

At the moment, an MTL model can be learnt using the mixup samples. It is expected that the high-level representation on the ASC branch encodes the acoustic scene features and the one on the AEC branch encodes acoustic event information. In this subsection, we will demonstrate that these high-level features can be utilized to improve the performance for the specific single task. The improving procedure includes two steps: reconstruction of the network and fine-tuning.

To reconstruct the network, the output layers of both branches are removed. Then a fully connected layer with 1024 neurons is added on top of each branch and they are connected onto a new $P$-dimensional softmax output layer for the single task. The final structure is shown in Fig. 4. Specifically, the outputs of the fully connected layers ($H_{16}^S$, $H_{16}^E$) are added up and fed into the output layer. It has:

$$H = g(W(H_{16}^S + H_{16}^E) + b) \qquad (6)$$

where $H$ denotes the output values for the newly-appended output layer; $W$ and $b$ are weights and bias; $g(\cdot)$ is the activation function.

The reconstructed network is then fine-tuned using the original single-task samples (rather than the mixup samples). Before fine-tuning, the softmax output layer and the fully connected layers are randomly initialized while the other layers are restored from the pre-trained MTL model. The output layer and the fully connected layers newly appended are fine-tuned, while the other layers are left frozen. In this way, this model is expected to be able to extract both acoustic scene and event features while fuse these features to get better performance for the specific single task.

### E. Late Fusion Ensemble

Late fusion ensemble is a commonly used strategy in the environmental sound analysis. In this letter, we also apply late fusion ensemble using the final network (Fig. 4). Specifically, by using different sample divisions, four models are obtained for the final network. The network is trained repeatedly three times. As mentioned above, the ASC spectrograms utilized are generated in a triply-generated manner (left, right, left+right). To get more different models, we train models using quadruple-generated ASC spectrograms (left, right, left+right, left-right) as well. As a result, 24 different classifiers are available. The output scores of the above classifiers for each class are accumulated. The class with maximum output scores is selected as the per-segment predicted result. Based on these results, per-recording results can be calculated using a mechanism like majority voting. Similar to other works, per-recording accuracy is chosen as a performance metric in this letter.

## III. EXPERIMENTS AND RESULTS

### A. Datasets

We perform ASC task on TUT acoustic scene 2017 dataset and AEC task on ESC-50 dataset. The TUT evaluation set is left as test set. The TUT development set is used as training set and validation set. Each time one subset in development set is selected as validation set and the other three subsets are used for training. As a result, all models are trained and tested four times. Averaged on the four results, a final accuracy is obtained for the models. To evaluate the models more accurately, this procedure is repeated three times for each model. Performances of the models are reported using the average accuracy and the standard deviation over the three trials. Five different divisions of the dataset were provided by the publisher of ESC-50. In each division, the recordings were divided into two parts. We use the small part as test set and the large one as training set and validation set. Similarly, the amount ratio of the latter two is 3:1. To match the division of TUT acoustic scene 2017, only the first four divisions are utilized in our experiments. In the process of mixup samples generation, spectrograms from training (validation) set of TUT development set are randomly mixed with the ones from training (validation) set of ESC-50.

## B. Parameter Setting

The experiments are implemented on the TensorFlow. The models are learnt by using a mini-batch size of 256 and an Adam optimizer with a learning rate of 0.0001. In normal training, the epoch number is 200 for ASC and 500 for AEC. During fine-tuning, it is set as 100 for ASC and 500 for AEC. The parameter $\alpha$ in equation 4 and 5 is chosen experimentally; it is set as 0.7 for ASC and 0.6 for AEC in the experiments.

TABLE I: Performance comparisons among different models

| Models | ASC Accuracy | AEC Accuracy |
|---|---|---|
| Baseline model for ASC | 73.91 ±0.74% | / |
| Baseline model for AEC | / | 63.5 ±0.39% |
| MTL model | 76.64 ±0.18% | 66.63 ±1.02% |
| MTL model with inter-connections | 77.70 ±0.52% | 68.50 ±0.65% |
| Fine-tuned model for ASC | **79.44 ±0.38%** | / |
| Fine-tuned model for AEC | / | **70.79 ±0.63%** |

## C. Effect of Mixup based MTL

In this subsection, it will be demonstrated that mixup based MTL can improve the performances of both tasks. As shown in Table Ⅰ, compared with baseline models, an improvement of 2.73 percent and an improvement of 3.13 percent are achieved respectively for ASC task and AEC task on a simple mixup based MTL model. The proposed mixup based MTL is low-cost yet effective for environmental sound analysis. Moreover, it is shown that adding inter-connections between the branches of the MTL model can further improve accuracy for both tasks. In particular, a 1.06 percent and a 1.87 percent improvements are achieved for the ASC and AEC tasks. This result is consistent with the supposition that acoustic scene and acoustic event are mutually reinforcing in classification.

## D. Effect of cross-task fusion

To take advantage of the complementary characteristic of ASC and AEC, the mixup MTL model is in turn modified into a single-task classification model. Most of the pre-trained parameters are preserved and both acoustic scene features and event features can serve for the single task using the method of fine-tuning. As can be seen in Table Ⅰ, the accuracy of the fine-tuned model for ASC is 79.44 percent, which is an improvement of 1.74 percent over the MTL model with inter-connections. Similarly, the accuracy of the fine-tuned model for AEC is 70.79 percent. Compared with the MTL model with inter-connections, there is an improvement of 2.29 percent.

## E. Effect of two-stage training

When the network shown in Fig. 4 is directly trained using the TUT acoustic scene 2017 dataset and the ESC-50 dataset respectively, an accuracy of 74.62 ±0.15 percent is achieved for ASC and 67.45 ±0.60 percent for AEC. Compared with the fine-tuned models, they are both declined. From this point of view, the two-stage method can improve performance effectively for both tasks by introducing additional dataset.

In conventional transfer learning, the model pre-trained by an auxiliary dataset is transferred into a main task by fine-tuning using the main task dataset. We also perform conventional transfer learning on the network shown in Fig. 4. The model is pre-trained using the ESC-50 dataset and fine-tuned using the TUT acoustic scene 2017 dataset. An accuracy of 72.54 ±1.52 percent is obtained for ASC. Using a similar method, an accuracy of 61.17 ±1.01 percent is achieved for AEC. As can be seen, the two-stage method has outperformed the conventional transfer learning method (at least in our case). One of the reasons is that the proposed method is able to guide the network to learn task-specific features and domain knowledge can be utilized in the fusion.

## F. Performance Evaluation

By applying late fusion ensemble (24 classifiers are used), an ASC accuracy of 84.50 percent and an AEC accuracy of 77.50 percent are achieved by our method. Note that the experiments in Table Ⅰ are conducted using only the triply-generated acoustic scene spectrograms. To evaluate the performance, state-of-the-art results on the TUT acoustic scene 2017 and ESC-50 datasets are listed in Table Ⅱ. Our method has achieved a satisfactory result on the ASC task, which is improved by 1.2 percent compared with the accuracy of the first place in DCASE Challenge 2017[22]. In [22] data augmentation is applied and eight classifiers are late fused. One may argue that too many classifiers have been used in our method. Consequently, we also perform another group of experiments, which uses only eight fine-tuned classifiers (4 from triply-generated ASC spectrogram models, 4 from quadruple-generated models). The average result of these ensemble performances is 84.2 percent (over 9 combinations). It means that the superiority of our method remains even less classifiers integrated. However, the accuracy on AEC task is relatively common compared with the state-of-the-art results [23][18] on ESC-50. We will further improve it in the future research.

TABLE Ⅱ: Performance compared with other works

| ASC Models | ASC Accuracy | AEC Models | AEC Accuracy |
|---|---|---|---|
| [22] | 83.3% | [23] | **86.5%** |
| [16] | 80.4% | [18] | 84.2% |
| [17] | 77.7% | [19] | 83.5% |
| Our method | **84.5%** | Our method | 77.5% |

## IV. Conclusion

The MTL models trained by the mixup samples have achieved promising accuracies in both acoustic scene and acoustic event classification tasks according to the experiments. It is believed that there is a hopeful application prospect for the mixup based MTL on the joint acoustic analysis tasks. The proposed two-stage method can be considered as a transfer learning solution which is different with the conventional method. The proposed method is able to guide the network to extract task-specific features and fuse these task-specific features through fine-tuning. The two-stage method has achieved better performance than the conventional transfer learning method in our experiments.

Future works include: 1) verifying the two-stage method on more combinations of acoustic datasets; 2) using multiple spectrograms in ensemble to improve the AEC performance; 3) extending the proposed method to other application domains.